\documentclass[11pt]{article}
\usepackage[preprint]{acl}
\usepackage{times}
\usepackage{latexsym}
\usepackage[T1]{fontenc}
\usepackage[utf8]{inputenc}
\usepackage{microtype}
\usepackage{inconsolata}
\usepackage{graphicx}
\usepackage{amsmath,amssymb}
\usepackage{booktabs}
\usepackage{multirow}
\usepackage{CJKutf8}
\usepackage{float}
\usepackage{enumitem}
\usepackage{amsthm}
\newtheorem{proposition}{Proposition}
\graphicspath{{figs/}}
\newcommand{\WL}{WL}
\newcommand{\best}[1]{\textbf{#1}}

\title{Integrate the Reward, Don't Sample It: Window Likelihood for Telling Interruptions from Backchannels in Full-Duplex Spoken Dialogue Models}

\author{Guoqiang Hu$^{1,2}$ \quad Lichao Zhang$^{2}$ \quad Yizhou Peng$^{1}$ \quad Yuhao Luo$^{2}$ \\
  \bfseries Haifeng Hu$^{2}$ \quad Zhongqian Sun$^{2}$ \quad Wei Yang$^{2}$ \quad Eng Siong Chng$^{1}$ \\
  $^{1}$Nanyang Technological University, Singapore \quad $^{2}$Tencent \\
  \texttt{guoqiang002@e.ntu.edu.sg}}

\begin{document}
\maketitle

\begin{abstract}
When a user talks over a full-duplex dialogue model, it must decide whether the user is interrupting (stop) or backchannelling (carry on). Post-training can improve this discrimination or change how readily it stops. In our GRPO runs the second dominates: no configuration lowers both error rates on more than one model, and two released checkpoints discriminate no better than their bases. We trace this partly to how the reward is estimated. The probability of stopping inside a short window is a known function of a few stop probabilities the policy emits, yet GRPO estimates it by sampling. A group of rollouts that all agree gives no gradient, even when all fail; the push on the stop threshold is strongest where the model is unsure, whichever outcome is right; and whole-rollout credit reaches decisions the reward never judges. Confining the credit to the window removes only that leakage. Window likelihood (\WL) samples the words but integrates over the stop decisions, so the probability of the correct outcome is computed exactly and its logarithm maximised. On HumDial-FDBench, across four open models, \WL\ gives the best F1 and lowers both error rates by more than their confidence intervals, without talking into silence.
\end{abstract}

\section{Introduction}

Full-duplex spoken dialogue models \citep{nguyen2023dgslm,defossez2024moshi,veluri2024syncllm,wang2025freezeomni,liu2026lychee} listen while they speak, so the user can talk over them at any moment. Each time this happens, the model has to decide, while the overlap is still going on, whether to \emph{yield} (stop speaking) or to \emph{hold} (keep speaking). A speaker who is being interrupted (``wait, what about the price?'') drops out; one who is only receiving a backchannel (``uh-huh''; \citealp{yngve1970edgewise}) carries on. The two events begin the same way, with the other voice on top of one's own, and call for opposite actions, so yielding or holding correctly is a discrimination problem that has to be solved early.

\begin{figure*}[t]
\centering
\includegraphics[width=\textwidth]{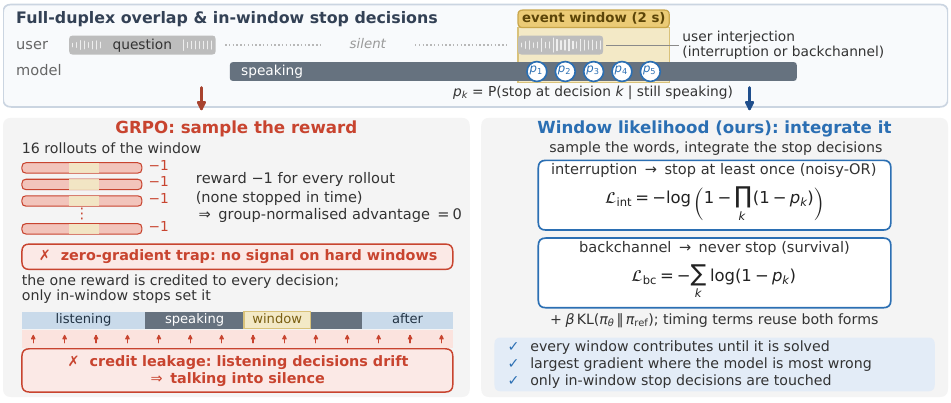}
\caption{One event window, two ways to learn from it. Top: the user interjects while the model is speaking; inside the 2-s window the model makes a few binary stop decisions, with stop probabilities $p_1,\ldots,p_K$. Left (GRPO): 16 sampled rollouts, one reward each (as $\pm1$), and that reward is spread over every decision in the rollout. Right (window likelihood): one sampled continuation; the expected reward is computed from its $p_k$ in closed form and maximised as a likelihood.}
\label{fig:overview}
\end{figure*}

The open models we tested get this wrong in both directions. Moshika yields to under half of real interruptions in our tests (Table~\ref{tab:main}; cf.\ \citealp{arora2025talkingturns}), while the released Lychee-FD \citep{liu2026lychee} stops on half of user backchannels. Post-training can make the model stop more or less readily; that trades one error for the other and, in signal-detection terms \citep{macmillan2005detection}, only moves the criterion $c$ (\S\ref{sec:setting}). Lowering both errors requires telling the two events apart better, a rise in sensitivity $d'$. In our runs, standard GRPO, which credits the whole rollout, never improves both at once. The GRPO checkpoints Kyutai released for PersonaPlex and Moshika \citep{ohashi2026multifaceted} keep the sensitivity of their base models, and our own GRPO runs move the criterion so far that two of four models almost never stop.

We argue that the way GRPO \citep{shao2024deepseekmath} estimates the window reward, whether the model stops within 2~s of the interjection, is partly to blame. The probability that the model stops inside a 2-s window is not an opaque quantity: the policy has already emitted the few stop probabilities that determine it (\S\ref{sec:setting}). GRPO nevertheless treats the reward as a black box and estimates it from a group of rollouts (Figure~\ref{fig:overview}), with three consequences. A group whose rollouts all receive the same reward has zero advantage, so the windows the model gets most wrong teach it nothing in that update (a \emph{zero-gradient trap}). GRPO's expected push on the stop logit is largest where the model is unsure, whichever outcome is correct (a \emph{criterion push}). And whole-rollout GRPO credits each sampled reward to every decision in the rollout, including listening decisions such as when to start talking (\emph{credit leakage}); after whole-rollout GRPO, Lychee-FD starts new, well-formed turns after its answer although the user says nothing (\S\ref{sec:setup}).

We therefore compute the expected reward instead of sampling it. \emph{Window likelihood} (\WL) still samples the words of a continuation per window, but computes, from its stop probabilities, the probability that the model stops in the window. It then maximises the log-probability of the correct outcome: a noisy-OR for interruptions (stop at least once) and a survival product for backchannels (never stop).

We post-train four open models that expose the stop decision in three different ways and evaluate them, together with two released GRPO checkpoints, on HumDial-FDBench \citep{zhao2026humdial,wang2026humdialtrack} in a closed loop without an LLM judge. On all four, \WL\ gains sensitivity while moving the criterion by at most 0.3, whereas whole-rollout GRPO moves the criterion by 1.6 to 2.4 and raises $d'$ only on the two models where it did not collapse (\S\ref{sec:results}). We arrange the baselines as a ladder (Table~\ref{tab:ladder}) that first restricts where GRPO's credit lands and then replaces the sampled reward by its closed form with the timing terms (\S\ref{sec:setup}). Our contributions are:
\begin{itemize}[leftmargin=*,itemsep=1pt,topsep=3pt,parsep=0pt]
\item window likelihood, which computes the window reward's expectation over the policy's own stop probabilities in closed form and maximises its logarithm (\S\ref{sec:setting}--\ref{sec:method});
\item an analysis of where GRPO falls short on this reward, and why, with masks that separate credit leakage from the rest (\S\ref{sec:properties}, \S\ref{sec:leakage}--\ref{sec:signal});
\item a closed-loop, judge-free evaluation of four post-trained models and two released GRPO checkpoints that separates sensitivity, criterion and silence behaviour (\S\ref{sec:exp}--\ref{sec:analysis}).
\end{itemize}

\section{Related Work}

Open full-duplex models \citep{nguyen2023dgslm,defossez2024moshi,veluri2024syncllm,zhang2024duplex,ma2025lslm,zhang2025omniflatten,yu2025salmonnomni} differ in how, and whether, they expose the stop decision; we post-train at least one model with each kind of exposed stop decision: a control branch, a token, or a VAD-gated state head (\S\ref{sec:setup}).

Turn-taking has been post-trained with RL before. \citet{chen2025rlduplex} balance barge-in against backchannel continuation with online RL on their own 1.1B model and synthetic tests. \citet{ohashi2026multifaceted} post-train Moshika and PersonaPlex with GRPO (16 rollouts per group), rule-based rewards on four axes and an LLM content reward; we evaluate their released checkpoints. DuplexPO \citep{li2026duplexpo} runs GRPO only inside critical windows, with the history teacher-forced, and does not evaluate holding through recorded user backchannels. ASPIRin \citep{hsiao2026aspirin} projects the action space onto speak/silent, and HiPLEX \citep{jung2026hiplex} routes sampled advantages to control decisions through event-causal masks; our GRPO speaking-only (\S\ref{sec:setup}) is a coarser relative of that mask. The group-based methods all sample the reward, so a uniform group gives no gradient even with window-local credit.

Uniform groups with zero-centred advantage are a known weakness of binary group rewards \citep{nie2026starvation}. The proposed fixes keep the sampler and repair it: a sign advantage \citep{nie2026starvation}, group splitting \citep{deng2026prism}, dynamic sampling \citep{yu2025dapo}. We do not compare to these repairs, which keep the sampled reward; our compute went to the credit masks of \S\ref{sec:setup}. Integrating over actions instead of sampling them is the idea behind expected policy gradients \citep{ciosek2018epg} and mean actor-critic \citep{allen2017mac}, and Rao-Blackwellisation more generally \citep{liu2019rao}; we apply it to a verifiable window reward whose expectation over the stop decisions is closed-form; \WL\ maximises its logarithm and is not an unbiased estimator of the policy gradient (Appendix~\ref{app:theory}). Dr.~GRPO \citep{liu2025drgrpo} analyses the reweighting that group normalisation induces; Proposition~\ref{prop:bias} is its two-outcome case, and dropping the normalisation gives the weight $S(1-S)$ in the window stop probability $S$ of \S\ref{sec:setting}, smaller still on hard windows (Figure~\ref{fig:signal}). Like on-policy distillation \citep{agarwal2024gkd,lu2025onpolicydistillation}, \WL\ scores the policy's own samples with a dense teacher-forced loss. Its target, though, is a rule on the window outcome, not a teacher's token distribution, and the interruption term is a set-level noisy-OR that does not say where to stop.

\section{Problem Formulation}
\label{sec:setting}

\paragraph{Stop decisions.} A full-duplex model reads the user's audio stream and writes its own output stream in real time. While it is speaking, it repeatedly decides whether to continue or to stop, at its own rate. We write the probability of stopping at decision $k$ as
\begin{equation}
p_k=\pi_\theta\big(\mathrm{stop}\mid x_{\le t_k},\,y_{<t_k}\big),
\label{eq:pk}
\end{equation}
where $\pi_\theta$ is the policy, $t_k$ the time of the decision, $x_{\le t_k}$ the user's audio so far, and $y_{<t_k}$ the model's output so far, which contains no stop. The four models we post-train expose the stop decision as a control branch (Lychee-FD), a token (MiniCPM-o, BayLing) or a state head behind a VAD gate (Freeze-Omni). The Moshi family (PersonaPlex, Moshika) exposes none; on them we train nothing and evaluate the base models and Kyutai's released GRPO checkpoints. While silent, the model likewise decides at each step whether to start speaking; we call these \emph{listening decisions}.

\paragraph{Events and reward.} A training example is an \emph{event}: the user asks a question, the model starts to answer, and at time $t_0$, while the model is speaking, the user \emph{interjects}. The interjection is labelled $\ell\in\{\mathrm{int},\mathrm{bc}\}$, an interruption or a backchannel. The model should stop within the \emph{event window} $W=[t_0,t_0+2\,\mathrm{s}]$ after an interruption and keep speaking through it after a backchannel. With $o=1$ if the model stops inside $W$ and $o=0$ otherwise, the window reward is
\begin{equation}
r=\begin{cases}o, & \ell=\mathrm{int},\\ 1-o, & \ell=\mathrm{bc},\end{cases}
\label{eq:r}
\end{equation}
and post-training maximises its expectation $J(\theta)=\mathbb{E}[r]$ over events and over the policy's behaviour.

\paragraph{Given the content, the expected reward is closed-form.} Let $u$ be the content the model would produce if it kept speaking through $W$, and let $p_1,\ldots,p_K$ be the stop probabilities at the $K$ decisions it reaches inside $W$. The probability that the model stops at one of these decisions is
\begin{equation}
S \;=\; 1-\prod_{k=1}^{K}(1-p_k),
\label{eq:S}
\end{equation}
with no independence assumption: each $p_k$ is a discrete-time hazard, already conditioned on the model not having stopped before $k$. The conditional expected reward $\mathbb{E}[r\mid u]$ is therefore $S$ for an interruption and $1-S$ for a backchannel.

\paragraph{Metrics.} The yield rate $I$ is the fraction of test interruptions after which the model stops inside $W$, and the false-stop rate $B$ the fraction of backchannels after which it does. We rank models by
\begin{equation}
\mathrm{F1}=\frac{2\,I\,(1-B)}{I+1-B},
\label{eq:f1}
\end{equation}
the harmonic mean of the yield rate $I$ and the hold rate $1-B$; we call it F1 although neither rate is a precision. Because $I$ and $B$ trade against each other, we also summarise each operating point by a sensitivity $d'=\Phi^{-1}(I)-\Phi^{-1}(B)$, with $\Phi$ the standard normal CDF, and a criterion (the stop threshold) $c=-\tfrac12[\Phi^{-1}(I)+\Phi^{-1}(B)]$ \citep{macmillan2005detection}, with a log-linear correction for extreme rates \citep{hautus1995corrections}. A larger $d'$ means the two events are told apart better; a positive $c$ means reluctance to stop, and $c=0$ means the miss rate $1-I$ equals the false-stop rate $B$. Under the equal-variance model, stopping more or less readily moves only $c$.

\section{Window Likelihood}
\label{sec:method}
\WL\ replaces GRPO's sampled estimate of the window reward (Figure~\ref{fig:overview}, left) with the closed form of Equation~(\ref{eq:S}) (Figure~\ref{fig:overview}, right). It still draws a continuation per window, but the window reward is computed, never sampled; the operative ingredient is the log-likelihood of that computed probability, not the integration alone (\S\ref{sec:properties}). An event term scores each window by the log-probability of the correct outcome; two timing terms of the same form and a KL term complete the loss (\S\ref{sec:objective}).

\subsection{Objective}
\label{sec:objective}
\WL\ maximises the log-probability of the correct outcome under (\ref{eq:S}),
\begin{align}
\mathcal{L}_{\mathrm{int}} &= -\log S = -\log\Big(1-\prod_{k}(1-p_k)\Big), \label{eq:int}\\
\mathcal{L}_{\mathrm{bc}}  &= -\log(1-S) = -\sum_{k}\log(1-p_k). \label{eq:bc}
\end{align}
Equation~(\ref{eq:int}) is a noisy-OR, satisfied by any one stop; Equation~(\ref{eq:bc}) is a survival likelihood, a per-decision cross-entropy toward continuing. Both forms are standard: the noisy-OR from multiple-instance learning (only the window is labelled, not a stop time; \citealp{maron1998mil,viola2006milboost}), the survival product from discrete-time survival analysis \citep{singer1993survival}; their use in place of a sampled turn-taking reward is new.

Per window, \WL\ draws one continuation from the current policy, stops included (two continuations on Lychee-FD). Let $j$ be its first stop inside $W$, or $K$ if it keeps speaking; $S$ is computed from $p_1,\ldots,p_j$ in a teacher-forced pass. If the continuation never stops, $\prod_k(1-p_k)$ is exactly the probability of holding given the content; otherwise $S$ is a lower bound, and the interruption term still raises $p_1,\ldots,p_j$. The continuation itself receives no gradient, and because the expectation is computed rather than estimated, no group and no advantage are needed (Appendix~\ref{app:theory}).

The same two forms also apply to timing events that every training dialogue already contains. A \emph{pre-event} term $\mathcal{L}_{\mathrm{pre}}$ applies the survival form to the stop decisions between the model's onset and $t_0$, so that the model is still speaking when the user comes in. A \emph{turn-taking} term $\mathcal{L}_{\mathrm{tt}}$ applies the noisy-OR to the start-speaking decisions in the 2~s after the user's first utterance ends, and the survival form to those decisions while the user is still talking. The total loss is
\begin{equation}
{\medmuskip=2mu\thickmuskip=3mu\mathcal{L}=\mathcal{L}_{\mathrm{event}}+0.5\,\mathcal{L}_{\mathrm{pre}}+\mathcal{L}_{\mathrm{tt}}+\beta\,\mathrm{KL}(\pi_\theta\|\pi_{\mathrm{ref}}),}
\label{eq:total}
\end{equation}
with $\mathcal{L}_{\mathrm{event}}$ equal to (\ref{eq:int}) or (\ref{eq:bc}) according to the label, $\pi_{\mathrm{ref}}$ the base model, and $\beta=0.05$ (terms used per model: Table~\ref{tab:budget}).

\subsection{Learning Signal Compared with GRPO}
\label{sec:properties}
Let $z_k$ be the logit of $p_k$. The objective differs from GRPO's sampled estimate in three ways.

\begin{figure}[t]
\centering
\includegraphics[width=0.72\columnwidth]{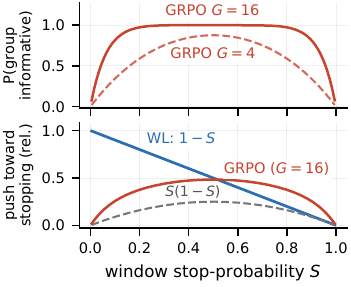}
\caption{Learning signal on one window as a function of its stop probability $S$. Top: probability that a GRPO group of $G$ rollouts is informative ($G{=}16$ solid, $G{=}4$ dashed). Bottom: weight of the push toward stopping on an interruption window (Proposition~\ref{prop:bias}, one decision) under \WL, exact $G{=}16$ GRPO, and the expected reward $S(1-S)$; backchannel windows mirror it; vertical scale arbitrary, only the shapes are comparable.}
\label{fig:signal}
\end{figure}

\paragraph{Every window contributes until it is solved (no zero-gradient trap).} A GRPO group of $G$ rollouts carries signal only if its rollouts disagree, which happens with probability $1-S^G-(1-S)^G$. This probability vanishes at both ends of Figure~\ref{fig:signal} (top): on solved windows, as it should, and on windows the model gets completely wrong. At $S=0.02$ a group of 16 is informative with probability 0.28. \WL\ has no group to degenerate, so its gradient vanishes only at the solved end. Integration alone would not achieve this; the logarithm does. The expected reward has $\partial S/\partial z_k=(1-S)\,p_k$, which goes to zero with $S$. Its logarithm has $\partial\mathcal{L}_{\mathrm{int}}/\partial z_k = -(1-S)\,p_k/S$, and because $S\le\sum_k p_k\le KS$ the summed magnitude stays between $1-S$ and $K(1-S)$, bounded away from zero. The backchannel term is simpler: $\partial\mathcal{L}_{\mathrm{bc}}/\partial z_k = p_k$.

\paragraph{The push on the criterion vanishes at balanced errors ($c=0$).}
\begin{proposition}\label{prop:bias}
Let every stop logit in a window share an offset $\delta$, with one decision per window and the event terms alone. On an interruption window with stop probability $S$, the descent direction on $\delta$ points toward stopping with weight $1-S$ under \WL, $\sqrt{S(1-S)}$ under a group-normalised $\pm1$ reward at large $G$, and $S(1-S)$ under the expected reward. On a backchannel window it points toward holding with weights $S$, $\sqrt{S(1-S)}$ and $S(1-S)$.
\end{proposition}
Equivalently, the three objectives ascend $\log S$, $2\arcsin\sqrt{S}$ and $S$ on an interruption window; Figure~\ref{fig:signal} (bottom) plots the exact $G{=}16$ GRPO weight alongside. GRPO's weight peaks at $S=0.5$ for both classes, so its push is strongest on uncertain windows and indifferent to which outcome is correct. \WL's weights are the error probabilities (the calibration property of the log loss). Its push therefore vanishes when summed misses on interruption windows equal summed false stops on backchannel windows. With equally frequent classes, as in our batches, that fixed point is $c=0$ under the proposition's assumptions; several decisions per window and the timing terms shift it (Appendix~\ref{app:theory}). The proposition concerns the criterion only; $d'$ rises through gradients that differ between windows.

\paragraph{The event terms touch only in-window stop decisions (no credit leakage in the loss).} Listening decisions (when to start the first answer, whether to speak when the user is silent) do not appear in them, so the event terms, unlike GRPO's objective, need no mask. Shared parameters can still move the listening decisions, which is why three of the four models add the turn-taking term (Appendix~\ref{app:train}).

\section{Experimental Setup}
\label{sec:exp}
\label{sec:setup}
\label{sec:protocol}
\begin{table}[t]
\centering\footnotesize
\setlength{\tabcolsep}{3pt}
\resizebox{\columnwidth}{!}{%
\begin{tabular}{@{}llll@{}}
\toprule
Configuration & Window reward & Credited decisions & Timing \\
\midrule
GRPO G16 & sampled, $G{=}16$ & all decisions of the rollout & -- \\
GRPO speaking-only & sampled, $G{=}16$ & speaking decisions & -- \\
GRPO window-only & sampled, $G{=}16$ & stop decisions inside $W$ & -- \\
\WL & computed, (\ref{eq:S}) & stop decisions inside $W$ & pre / tt \\
\bottomrule
\end{tabular}}
\caption{Configurations trained per model, read top to bottom as a ladder (\S\ref{sec:setup}). The step from GRPO window-only to \WL\ changes both how the window reward enters the loss and the timing terms; at the same budget it also changes the number of distinct windows and, on MiniCPM-o, of updates (Appendix~\ref{app:train}), which we do not control for. On BayLing, GRPO credits all text tokens. Timing: pre-event (pre) and turn-taking (tt) terms (\S\ref{sec:objective}; per model in Table~\ref{tab:budget}).}
\label{tab:ladder}
\end{table}

\begin{table*}[t]
\centering\footnotesize
\setlength{\tabcolsep}{3.4pt}
\begin{tabular}{@{}llrrrrrrl@{}}
\toprule
Model & Configuration & $I$ (\%) $\uparrow$ & $B$ (\%) $\downarrow$ & F1 $\uparrow$ & $d'$ $\uparrow$ & $c$ & Lat.\ (s) & Re-open / speech \\
\midrule
\multirow{5}{*}{Lychee-FD (zh)}
 & Base & 85.3 {\scriptsize$\pm$1.9} & 50.5 {\scriptsize$\pm$5.9} & 62.6 {\scriptsize$\pm$4.8} & 1.04 & $-$0.53 & 0.40 & 2.5\% / 0.4~s \\
 & GRPO G16 & 61.3 {\scriptsize$\pm$2.3} & \best{0.3} & 75.9 {\scriptsize$\pm$1.8} & 2.86$^\circ$ & 1.14 & 1.20 & 100\% / 45.8~s \\
 & GRPO speaking-only & 73.1 {\scriptsize$\pm$2.1} & 0.7 & 84.2 {\scriptsize$\pm$1.4} & \best{3.01}$^\circ$ & 0.89 & 1.20 & 0\% / 0 s \\
 & GRPO window-only & 66.8 {\scriptsize$\pm$2.2} & 0.7 & 79.9 {\scriptsize$\pm$1.6} & 2.83$^\circ$ & 0.98 & 1.20 & 0\% / 0 s \\
 & \WL\ (ours) & \best{95.9 {\scriptsize$\pm$1.0}} & 12.8 {\scriptsize$\pm$3.9} & \best{91.4 {\scriptsize$\pm$2.1}} & 2.87 & $-$0.30 & 0.40 & 0\% / 0 s \\
\midrule
\multirow{5}{*}{MiniCPM-o (zh)}
 & Base & 16.5 {\scriptsize$\pm$1.9} & 10.0 {\scriptsize$\pm$3.3} & 27.9 {\scriptsize$\pm$2.7} & 0.30 & 1.12 & 0.92 & 0\% / 0 s \\
 & GRPO G16 & \best{92.1 {\scriptsize$\pm$1.4}} & 34.7 {\scriptsize$\pm$5.4} & 76.4 {\scriptsize$\pm$3.8} & 1.80 & $-$0.51 & 1.08 & 0\% / 0 s \\
 & GRPO speaking-only & 80.1 {\scriptsize$\pm$1.9} & 11.9 {\scriptsize$\pm$3.6} & 83.9 {\scriptsize$\pm$2.0} & 2.02 & 0.17 & 1.31 & 0\% / 0 s \\
 & GRPO window-only & 66.9 {\scriptsize$\pm$2.3} & 3.4 {\scriptsize$\pm$2.1} & 79.0 {\scriptsize$\pm$1.7} & 2.24 & 0.68 & 1.40 & 0\% / 0 s \\
 & \WL\ (ours) & 81.3 {\scriptsize$\pm$1.9} & \best{0.3} & \best{89.6 {\scriptsize$\pm$1.2}} & \best{3.46}$^\circ$ & 0.84 & 1.36 & 0\% / 0 s \\
\midrule
\multirow{5}{*}{BayLing (en)}
 & Base & 76.0 {\scriptsize$\pm$2.1} & 79.5 {\scriptsize$\pm$4.5} & 32.3 {\scriptsize$\pm$5.6} & $-$0.12 & $-$0.76 & 0.80 & 100\% / 69.0~s \\
 & GRPO G16 & 4.9 {\scriptsize$\pm$1.1} & 4.7 {\scriptsize$\pm$2.5} & 9.4 {\scriptsize$\pm$2.0} & 0.01 & 1.65 & 1.38 & 100\% / 67.9~s \\
 & GRPO speaking-only & 77.7 {\scriptsize$\pm$2.0} & 79.3 {\scriptsize$\pm$4.7} & 32.7 {\scriptsize$\pm$5.8} & $-$0.05 & $-$0.79 & 0.80 & 100\% / 66.6~s \\
 & GRPO window-only & 51.0 {\scriptsize$\pm$2.6} & 51.6 {\scriptsize$\pm$5.6} & 49.7 {\scriptsize$\pm$3.3} & $-$0.01 & $-$0.03 & 1.30 & 100\% / 71.4~s \\
 & \WL\ (ours) & \best{95.7 {\scriptsize$\pm$1.0}} & \best{22.4 {\scriptsize$\pm$4.8}} & \best{85.7 {\scriptsize$\pm$2.9}} & \best{2.46} & $-$0.48 & 0.80 & 0\% / 0 s \\
\midrule
\multirow{5}{*}{Freeze-Omni (zh)}
 & Base & 11.7 {\scriptsize$\pm$1.5} & 37.6 {\scriptsize$\pm$5.4} & 19.7 {\scriptsize$\pm$2.2} & $-$0.88 & 0.75 & 1.68 & 0\% / 0 s \\
 & GRPO G16 & 0.3 & 0.0 & 0.5 & $-$0.06$^\ast$ & 2.62 & -- & 0\% / 0 s \\
 & GRPO speaking-only & 0.4 & 0.0 & 0.8 & 0.14$^\ast$ & 2.68 & 2.00 & 0\% / 0 s \\
 & GRPO window-only & 1.0 & 0.0 & 1.9 & $-$0.10$^\ast$ & 2.14 & -- & 0\% / 0 s$^\ddagger$ \\
 & \WL\ (ours) & \best{26.0 {\scriptsize$\pm$2.2}} & \best{9.1 {\scriptsize$\pm$3.2}} & \best{40.5 {\scriptsize$\pm$2.7}} & \best{0.69} & 0.99 & 0.46 & 0\% / 0 s \\
\midrule
\multirow{2}{*}{PersonaPlex (en)}
 & Base & 93.8 {\scriptsize$\pm$1.2} & 54.0 {\scriptsize$\pm$5.8} & 61.7 {\scriptsize$\pm$5.4} & 1.44 & $-$0.82 & 0.64 & 10\% / 1.4~s \\
 & Kyutai GRPO$^\dagger$ & 96.0 {\scriptsize$\pm$1.0} & 63.9 {\scriptsize$\pm$5.7} & 52.4 {\scriptsize$\pm$6.0} & 1.39 & $-$1.05 & 0.56 & 100\% / 40.9~s \\
\midrule
\multirow{2}{*}{Moshika (en)}
 & Base & 47.2 {\scriptsize$\pm$2.7} & 27.1 {\scriptsize$\pm$5.9} & 57.3 {\scriptsize$\pm$2.8} & 0.53 & 0.34 & 1.36 & 94.6\% / 13.4~s \\
 & Kyutai GRPO$^\dagger$ & 59.4 {\scriptsize$\pm$3.0} & 40.7 {\scriptsize$\pm$6.5} & 59.4 {\scriptsize$\pm$3.6} & 0.47 & 0.00 & 1.14 & 100\% / 52.4~s \\
\bottomrule

\end{tabular}
\caption{HumDial-FDBench, full test set, the same 2-s window for all models. $I$: yield rate on interruptions; $B$: false-stop rate on backchannels; F1 as in (\ref{eq:f1}); $d'$, $c$ as in \S\ref{sec:setting} (positive $c$: reluctant to stop); Lat.: median yield latency (--: a single yield); Re-open / speech: re-open rate and seconds of speech in the 100-s silence test (\S\ref{sec:setup}). $\pm$: 95\% bootstrap half-width over test items, omitted at or near 0/100\%. $^\circ$rests on a rate below 1\%; $^\ast$both rates near zero. Freeze-Omni GRPO rows rest on 375, 779 and 104 valid interruptions; $^\ddagger$six dialogues answered. $^\dagger$Kyutai's released checkpoints \citep{ohashi2026multifaceted}, evaluated only. Bold: best of $I$, $B$, F1 and $d'$ within a trained model, excluding collapsed runs (yield below 5\%). \WL\ rows are the first seed; seeds, valid counts and CIs for $d'$, $c$: Appendices~\ref{app:train}, \ref{app:results}.}
\label{tab:main}
\end{table*}
\paragraph{Models and data.} We post-train four open models (Appendix~\ref{app:train}): the control head of Lychee-FD, attention LoRAs on MiniCPM-o~4.5 \citep{cui2026minicpmo} and BayLing-Duplex \citep{fang2026bayling}, and the state head of Freeze-Omni. All are trained only on the interruption and backchannel clips of the HumDial Track-2 training set \citep{wang2026humdialtrack}, in the model's test language; we construct no silence or synthetic data, and training and test speakers and dialogues are disjoint.

\paragraph{Baselines and masks.} Table~\ref{tab:ladder} orders the configurations as a ladder. Reading down, the first two steps change only where the credit lands; the last replaces the sampled reward by its closed form and adds the timing terms. \emph{GRPO G16} is standard whole-rollout GRPO: 16 rollouts of one window per step, the window reward (\ref{eq:r}) as $\pm1$, and the group-normalised advantage applied to every decision of the rollout. It trains the same parameters as \WL\ and reuses \WL's learning rate and KL weight without retuning; each model gets one run. It draws as many event-window samples as \WL\ draws continuations, so it sees 8 to 16 times fewer distinct windows (Appendix~\ref{app:train}). Two \emph{masks} keep GRPO's reward, data and budget and change only where the advantage lands. \emph{Speaking-only} credits decisions made while the model speaks; \emph{window-only} credits only those inside the event window, the window-local credit of DuplexPO \citep{li2026duplexpo} without its teacher-forced history. On PersonaPlex and Moshika \citep{roy2026personaplex,defossez2024moshi}, which expose no stop decision, we evaluate the base models and Kyutai's released GRPO checkpoints \citep{ohashi2026multifaceted}, trained on 4,000~h of dialogue with rule-based and LLM-judged rewards.

\paragraph{Test set and closed loop.} HumDial-FDBench \citep{zhao2026humdial} is, to our knowledge, the only turn-taking benchmark with real-recorded interruptions and user backchannels in two languages: 1,500 interruptions in five categories and 300 backchannels per language (Table~\ref{tab:cats}, Appendix~\ref{app:eval}). We test Lychee-FD, MiniCPM-o and Freeze-Omni in Chinese and the others in English. A fixed user track cannot guarantee that an interjection arrives while the model is speaking, so we evaluate all models in a closed loop. The simulator waits for the model to start answering and injects the recorded interjection 0.8, 1.6 or 2.4~s later. Items where the model has already stopped are dropped, so $I$ and $B$ are rates over valid sets whose size varies by configuration (counts in Appendix~\ref{app:results}). For the Moshi family, which has no stop decision, a stop is a gap of at least 0.8~s in output energy.

\paragraph{Readout.} Besides $I$, $B$, F1, $d'$ and $c$, we report the median latency of yields. We use no LLM judge: transcript judges agree only moderately with humans on turn-taking (Krippendorff's $\alpha=0.61$, \citealp{lin2026fdb2}; below 80\% agreement, \citealp{arora2025talkingturns}).

\paragraph{Silence test.} In each of 40 dialogues (English, the same for all six models) the user asks one question and then stays silent for 100~s. We report how often the model re-opens the floor, that is, talks into silence, and for how many seconds (Appendix~\ref{app:eval}).

\section{Results and Analysis}
\label{sec:analysis}

\begin{figure*}[!tb]
\centering
\includegraphics[width=\textwidth]{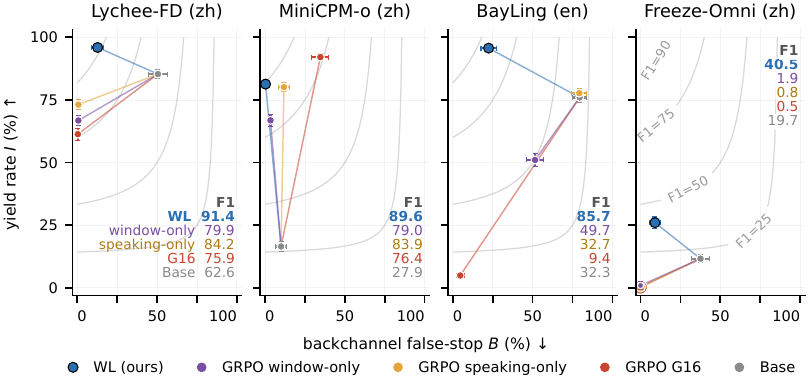}
\caption{Operating points on HumDial-FDBench: yield rate $I$ against false-stop rate $B$ (95\% CIs; F1 iso-lines at 25, 50, 75, 90; the ideal is top-left). Lines join each configuration to its base; the corner lists give F1 in the order named in the first panel. GRPO configurations mostly move along the trade-off (up-right or down-left); the exception is window-only on MiniCPM-o (up-left); speaking-only there moves almost straight up. \WL\ moves toward the corner on all four models. PersonaPlex and Moshika, evaluated only, are in Table~\ref{tab:main}.}
\label{fig:operating}
\end{figure*}
\begin{figure*}[!tb]
\centering
\includegraphics[width=\textwidth]{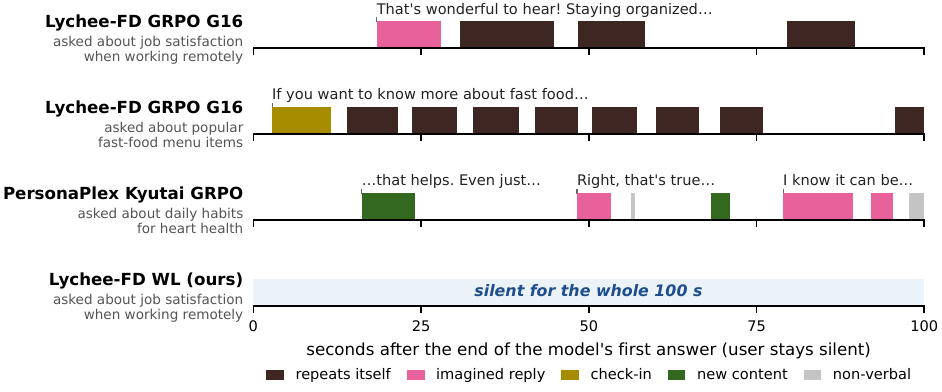}
\caption{Four dialogues from the 100-s silence test; each bar is a turn the model starts on its own, coloured by category; selected turns show their opening words. After GRPO G16, Lychee-FD answers an imagined user turn and then repeats itself, or says one check-in nine times (coloured as repeats after the first); Kyutai's PersonaPlex checkpoint plays both sides of a conversation; \WL\ stays silent. Table~\ref{tab:silence_text} summarises all 40 dialogues per configuration.}
\label{fig:selftalk}
\end{figure*}
\begin{figure*}[!t]
\centering
\includegraphics[width=\textwidth]{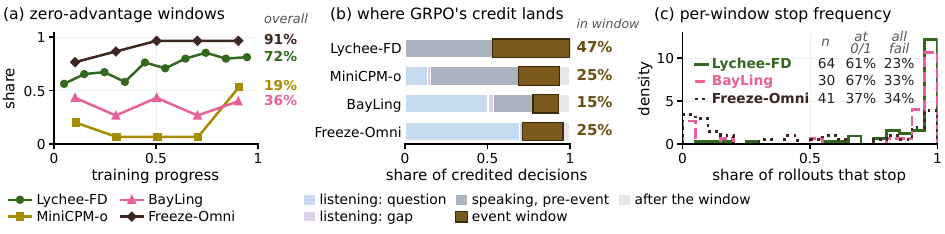}
\caption{Where the sampled signal goes. (a) Share of training windows with zero advantage (valid rollouts all with the same reward, or fewer than two valid rollouts) over training; overall share at the line ends. (b) Share of credited decisions by dialogue state on base-policy rollouts (truncated 2.5~s after injection, so post-stop listening decisions are not shown); only the dark segment, whose share is on the right, lies inside the window. (c) Per-window stop frequency among 16 base-policy rollouts (MiniCPM-o kept no per-window log); the table gives the number of windows $n$, the share at 0 or 1, where the group is uniform, and the share on which all 16 rollouts fail by label.}
\label{fig:diag}
\end{figure*}
\subsection{\WL\ Lowers Both Errors on Every Model; GRPO on at Most One}
\label{sec:results}
\WL\ has the best F1 on all four post-trained models, with 95\% intervals that do not overlap the runner-up's (Table~\ref{tab:main}). On Lychee-FD and MiniCPM-o, where GRPO trained stably, it leads the best GRPO configuration by 7.2 and 5.7 F1; GRPO G16 collapsed on BayLing, stopping on only 4.9\% of interruptions, and all three GRPO configurations collapsed on Freeze-Omni.

The gain is in discrimination. Moving the criterion along a fixed ROC curve can raise $I$ or lower $B$, never both. \WL\ does both, beyond the confidence intervals, on all four models (Figure~\ref{fig:operating}). Its $d'$ rises by 1.6 to 3.2 while $c$ moves by at most 0.3, toward 0 on three of them (Appendix~\ref{app:results}).

Table~\ref{tab:main} can be read in the order of Table~\ref{tab:ladder}. Whole-rollout credit (GRPO G16) swings $c$ by 1.6 to 2.4 on every model and lowers both errors on none; on Lychee-FD it reaches \WL's $d'$ (2.86 against 2.87), so under the equal-variance model the whole F1 gap there comes from the criterion. Restricting the credit to speaking decisions, then to the window, removes the talking into silence that GRPO G16 induced on Lychee-FD, and BayLing's collapse (\S\ref{sec:leakage}) but not the criterion move: GRPO window-only still moves $c$ by 1.5 on Lychee-FD and lowers both errors only on MiniCPM-o. The last step, from the sampled reward to its closed form with the timing terms (window-only to \WL), is the one that lowers both errors on every model: \WL\ leads GRPO window-only by 10.6 to 36.0 F1 wherever the latter does not collapse.

\subsection{Credit Leakage: Masks Remove It and the Re-opening It Caused}
\label{sec:leakage}
Whole-rollout GRPO changes behaviour the window reward never judges. After GRPO G16, Lychee-FD talks into every 100-s silence, as do Kyutai's checkpoints (evaluated, not trained) on PersonaPlex and Moshika (Table~\ref{tab:main}, last column; Figure~\ref{fig:raster}). Figure~\ref{fig:selftalk} shows what is said: the model answers an imagined user turn, repeats itself or carries on with new content; at most 22\% of the turns are check-ins (Table~\ref{tab:silence_text}). \WL\ adds no talking into silence on any model; its turn-taking term supervises listening decisions directly on three of them, so the masks, which add no such term, are the cleaner test of GRPO's mechanism.

Where GRPO's credit lands supports the leakage account. On base-policy rollouts, only 15 to 47\% of the decisions GRPO credits lie inside the event window (Figure~\ref{fig:diag}b). The rest fall before the event, on Freeze-Omni mostly while the user is still asking. On Lychee-FD, whose rollouts begin mid-answer, the credit outside the window in Figure~\ref{fig:diag}b is pre-event speech; beyond the 2.5-s truncation it falls on the listening decisions after the model's stop (Appendix~\ref{app:results}), a plausible route to the talking into silence. Masking that credit, by applying the advantage only while the model speaks and adding no timing term, removes Lychee-FD's re-opening, cuts MiniCPM-o's false stops by two thirds and keeps BayLing from collapsing (Table~\ref{tab:main}). Masking therefore removes the leakage and the talking into silence it caused on Lychee-FD; the criterion still moves (\S\ref{sec:signal}).

\subsection{The Trap and the Push Survive the Masks}
\label{sec:signal}
The zero-gradient trap survives every mask, because it is a property of the group, not of the credited positions, and it is common. On Lychee-FD, all valid rollouts agree on 72\% of training windows, and the share rises over the run (Figure~\ref{fig:diag}a; the zero-advantage share of the other models ranges from 19\% to 91\%). Under the base policy 61\% of Lychee-FD's logged windows are uniform (Figure~\ref{fig:diag}c), and by label all 16 rollouts fail on 23 to 34\% of them (Figure~\ref{fig:diag}c). The trap slows learning without stopping it. On MiniCPM-o, where few groups are uniform, GRPO speaking-only does gain sensitivity; on Lychee-FD, where most are, a criterion move alone can account for its $d'$ (Appendix~\ref{app:results}). MiniCPM-o is also where all three shortcomings are weakest, yet \WL\ still leads GRPO window-only by 10.6 F1; the uncontrolled differences in windows and updates (Table~\ref{tab:ladder}) may carry part of that gap.

The push is harder to pin down. Masking leaves Lychee-FD's swing ($c$ from $-0.53$ to 0.89, or to 0.98 under the window mask), and the $G{=}16$ push of Proposition~\ref{prop:bias}, summed over Lychee-FD's base-policy windows, points the same way (Appendix~\ref{app:theory}). On BayLing the net push is near zero; GRPO speaking-only indeed leaves $c$ unchanged, but window-only moves it to $-0.03$, which the sum does not predict. Freeze-Omni, collapsed in every GRPO run, is the exception (Appendix~\ref{app:theory}). The sign check thus succeeds on Lychee-FD, is mixed on BayLing and fails on Freeze-Omni, while the swing itself survives every mask. \WL\ has neither weakness. Every window in which the model is still speaking at the interjection contributes a gradient, and under Proposition~\ref{prop:bias}'s assumptions the net push vanishes at balanced errors.

\section{Conclusion}
\label{sec:discussion}
When the reward asks whether the policy stops within a window, it is a closed-form function of the policy's stop probabilities and can be integrated rather than sampled. Replacing the sampled reward by this closed form, with the timing terms, raised sensitivity on all four models we trained and left the criterion near its start, so both error rates fell, without any talking into silence. No GRPO configuration lowered both at once on more than one model. Whole-rollout GRPO credit moved the criterion and could make a model talk into silence; window-local credit still trailed \WL.

\section*{Limitations}
\label{sec:limitations}
\WL\ cannot change the interface through which the stop decision is exposed: Freeze-Omni's VAD-gated head cannot tell ``still asking'' from ``just asked'', so its F1 rises only from 19.7 to 40.5. Models without an explicit stop decision, such as the Moshi family, need a stop to be defined before \WL\ applies; we leave them to future work.

Our numbers are not comparable with the official HumDial leaderboard, which judges transcripts (\S\ref{sec:protocol}). The study uses one benchmark and one training seed per configuration except \WL\ on Lychee-FD (three) and MiniCPM-o (two). Window-only GRPO borrows DuplexPO's window-local credit but not its reward, data or teacher-forced history, so it is not a reproduction of DuplexPO. GRPO reuses \WL's learning rate and KL weight untuned. The roles of integration and the logarithm are argued (\S\ref{sec:properties}), not ablated, and \WL's $d'$ gain is not separated from its larger number of windows or its timing terms. The readout does not check what the model says, and the LoRA on MiniCPM-o and BayLing can change it; $d'$ and $c$ rest on an untested equal-variance model.

\bibliography{refs}

\appendix
\onecolumn

\section{Training Details}
\label{app:train}
\begin{table}[H]
\centering\footnotesize
\setlength{\tabcolsep}{3.4pt}
\resizebox{\ifdim\width>\textwidth\textwidth\else\width\fi}{!}{%
\begin{tabular}{@{}llllllr@{}}
\toprule
Model & Lang. & Stop decision & Trained part & Timing terms & Windows: \WL\ / GRPO & Event rollouts \\
\midrule
Lychee-FD \citeyearpar{liu2026lychee} & zh & control branch, one state per 0.4~s & control head & pre-event, turn-taking & 4,432 / 554 & 8,864 \\
MiniCPM-o 4.5 \citeyearpar{cui2026minicpmo} & zh & listen / eos tokens (text vocabulary) & LoRA (attention) & pre-event & 1,200 / 75 & 1,200 \\
BayLing-Duplex \citeyearpar{fang2026bayling} & en & [EPAD] / assistant tokens (text channel) & LoRA (attention) & pre-event, turn-taking & 2,400 / 150 & 2,400 \\
Freeze-Omni \citeyearpar{wang2025freezeomni} & zh & VAD gate + state head & state head & turn-taking & 2,400 / 150 & 2,400 \\
\bottomrule
\end{tabular}}
\caption{The four post-trained models and the training budget. \WL\ and GRPO draw the same number of event-window rollouts (``Event rollouts'': sampled continuations on event windows); GRPO spends 16 rollouts on each window, so it sees fewer distinct windows (``Windows''; the turn-taking term on Lychee-FD adds two windows per step).}
\label{tab:budget}
\end{table}
\paragraph{Common to all models.} The continuation scored by \WL\ is the model's own answer with the interjection injected as at test time, generated as in a GRPO rollout and re-scored with a teacher-forced forward pass. ``Event rollouts'' in Table~\ref{tab:budget} counts these continuations; GRPO draws the same number, 16 per window on one window per step, interruption and backchannel windows alternating, no resampling. GRPO mechanics: rollouts in which the model is not speaking at injection are dropped from the group rather than scored; the advantage is $(r-\mathrm{mean})/\mathrm{std}$ over the group and multiplies the mean log-probability of the rollout's decisions (all text positions for BayLing, control decisions otherwise); one update per group. The reward is not written in terms of listening decisions but depends on them through whether the model is speaking at injection.
\paragraph{Lychee-FD.} Control head only; 554 steps; learning rate $5\times10^{-6}$; KL weight 0.05; sampling temperature 1.0 on the control branch. \WL: 8 event windows $\times$ 2 continuations per step, plus 2 turn-taking windows; pre-event weight 0.5. GRPO: 1 window $\times$ 16 rollouts per step, $\pm1$ reward; the speaking-only configuration applies the advantage to decisions in the speaking state only. Lychee-FD base: \citet{liu2026lychee}, built on Step-Audio~2 \citep{wu2025stepaudio2}.
\paragraph{MiniCPM-o 4.5.} LoRA \citep{hu2021lora} on attention, rank 16, $\alpha=32$; learning rate $10^{-4}$; KL 0.05; pre-event weight 0.5, no turn-taking term; \WL\ 150 steps $\times$ 8 windows; GRPO 75 steps $\times$ 16 rollouts (1,200 rollouts each, so twice as many updates for \WL).
\paragraph{BayLing-Duplex.} LoRA on attention, rank 16, $\alpha=32$; learning rate $10^{-4}$; KL 0.05 over all text positions; pre-event weight 0.5; turn-taking term; 150 steps, 16 windows or rollouts per step.
\paragraph{Freeze-Omni.} State head only; learning rate $10^{-4}$; KL 0.05; turn-taking term (start within 2~s after the user's question ends; do not start while the user is still asking); 150 steps, 16 windows or rollouts per step.
\paragraph{Baseline details.} GRPO rollouts start before the model's first answer, or on Lychee-FD 1.5 to 2~s before the interjection, so the policy also makes listening decisions. GRPO has no timing terms and, on MiniCPM-o, takes half as many updates (Table~\ref{tab:budget}). Speaking-only GRPO still credits speaking decisions before the event; GRPO window-only credits only speaking decisions inside the event window, with the steps, learning rate and KL weight of GRPO G16, so each update reaches fewer positions and, since the advantage multiplies a mean log-probability, moves each of them more.
\paragraph{Seeds.} Two further training seeds of Lychee-FD \WL\ (identical configuration, run seeds 1 and 2) give $I$ = 95.5 / 95.3\%, $B$ = 8.7 / 8.8\% and F1 = 93.4 / 93.2 on the full test set, against 95.9\%, 12.8\% and 91.4 for the seed in Table~\ref{tab:main}. A second seed of MiniCPM-o \WL\ gives $I$ = 79.6\%, $B$ = 1.3\% and F1 = 88.1, against 81.3\%, 0.3\% and 89.6. None of the three re-opens in the silence test (0 of 40 dialogues each).

\section{Evaluation Details}
\label{app:eval}

\paragraph{Silence test.} 40 English HumDial questions, used for all six models (all answer English) so that the column is comparable across models. The model answers; after its first answer ends (energy-based end), 100~s of silence follow. Re-open: at least one new speech segment starts in the window. Seconds: total speech in the window. Both are computed over the dialogues that enter the window; a dialogue in which the model gives no answer cannot enter (Freeze-Omni GRPO G16: 4 of 40, window-only: 34 of 40; BayLing GRPO window-only: 1 of 40; Moshika base: 3 of 40).

\paragraph{Bootstrap.} Interruptions are resampled with replacement within each of the five categories and backchannels separately; $I$, $B$ and F1 are recomputed per resample. The 40 silence dialogues are resampled as a whole. 10,000 resamples, seed 0; $\pm=(q_{97.5}-q_{2.5})/2$. Cells at 0 or 100\%, and cells whose symmetric $\pm$ would cross 0 or 100\%, are shown without $\pm$; for a 0/40 cell the exact 95\% upper bound is 8.8\%.

\paragraph{Moshi-family readout.} Within-answer pauses exceed 0.8~s in only 1--2.5\% of cases, and a 1.2-s gap threshold preserves every ordering in Table~\ref{tab:main}.

\begin{table}[H]
\centering\small
\setlength{\tabcolsep}{3pt}
\begin{CJK*}{UTF8}{gbsn}
\begin{tabular}{@{}lp{6cm}p{6cm}@{}}
\toprule
Category & zh example & en example \\
\midrule
ask & 那怎么判断蛋白是否打发到位了呢？ & How often should these video check-ins ideally be scheduled? \\
deny & 不，这方法太老套了。 & That's not what I had in mind. \\
repeat & 稍等一下，刚才没听全，能重新说一下吗？ & Sorry, could you repeat that part about \ldots? \\
shift & 先不说这个了，我想问问\ldots & Actually, I'd like to change the subject. \\
wait & 外面有声音先别讲 & I'm busy, talk later. \\
backchannel & 嗯嗯 / 好的 / 是呀 & uh-huh / right \\
\bottomrule
\end{tabular}
\end{CJK*}
\caption{HumDial-FDBench interjection categories with examples. Each language has 300 interruptions per category and 300 backchannels.}
\label{tab:cats}
\end{table}

\section{Window Likelihood and the Sampled Reward}
\label{app:theory}
\paragraph{\WL\ and the expected reward.} Because the sampled continuation is held fixed, words and stops alike, \WL\ follows the gradient of $\log\mathbb{E}[r\mid u]$ for a fixed sample of the content $u$. This differs from ascending $J(\theta)=\mathbb{E}_u\,\mathbb{E}[r\mid u]$ in three ways: (i)~the logarithm sits inside the expectation; (ii)~the gradients through the sampled content and the sampled stop time are dropped; on Lychee-FD and Freeze-Omni the content does not depend on the trained parameters, but the stop time does; (iii)~after an early stop $S$ is truncated to a lower bound, so $1-S$ over-estimates the backchannel reward. Proposition~\ref{prop:bias} covers one decision per window and the event terms only. With $K$ decisions the \WL\ weights become $(1-S)R$ and $SR$, with $R=\sum_k p_k/S\in[1,K]$, and the pre-event and KL terms also act on the offset, so we do not derive where training settles.
\paragraph{Derivation of Proposition~\ref{prop:bias}.} With one decision of stop probability $S=\sigma(z+\delta)$, $\partial(-\log S)/\partial \delta=-(1-S)$ and $\partial(-\log(1-S))/\partial \delta=S$. The expected reward (\ref{eq:r}) of an interruption window is $S$, with gradient $S(1-S)$ (twice that for a $\pm1$ reward, a scale that group normalisation removes). For a group-normalised advantage with population mean $S$ and standard deviation $\sqrt{S(1-S)}$, the stopping rollouts carry $A=\sqrt{(1-S)/S}$ and score $\partial\log\pi/\partial z=1-S$, the holding ones $A=-\sqrt{S/(1-S)}$ and score $-S$, so $\mathbb{E}[A\,\partial\log\pi/\partial z]=\sqrt{S(1-S)}$. Backchannel windows flip the sign of the reward.

\paragraph{A check on the logged windows.} Summing the exact $G{=}16$ GRPO push of Proposition~\ref{prop:bias} over the logged training windows of Figure~\ref{fig:diag}c (per-window $S$ under the base policy) gives a net push toward holding on Lychee-FD ($-0.08$ per interruption--backchannel pair), the direction of its observed criterion shift. On BayLing the net push is near zero ($-0.01$), and the speaking-only run indeed leaves $c$ essentially unchanged ($-0.79$ against $-0.76$), so GRPO G16's swing comes from leakage. On Freeze-Omni the push points toward stopping ($+0.06$), opposite to the observed collapse, which also survives GRPO speaking-only; neither account explains it.

\section{Additional Results}
\label{app:results}
\begin{table}[H]
\centering\footnotesize\renewcommand{\arraystretch}{0.85}
\setlength{\tabcolsep}{3pt}
\begin{tabular}[t]{@{}lrrrrr@{}}
\toprule
Configuration & ask & deny & repeat & shift & wait \\
\midrule
\multicolumn{6}{@{}l}{\textit{Lychee-FD}} \\
\quad Base & 88.4 & 79.5 & 85.1 & 91.0 & 82.4 \\
\quad GRPO G16 & 78.0 & 45.3 & 66.0 & 81.0 & 36.3 \\
\quad GRPO speaking-only & 92.3 & 58.0 & 79.7 & 88.7 & 46.8 \\
\quad GRPO window-only & 85.7 & 53.7 & 70.0 & 86.3 & 38.3 \\
\quad \WL & 100.0 & 89.1 & 99.0 & 99.7 & 91.9 \\
\addlinespace[2pt]
\multicolumn{6}{@{}l}{\textit{MiniCPM-o}} \\
\quad Base & 19.4 & 14.9 & 13.4 & 19.8 & 14.8 \\
\quad GRPO G16 & 97.6 & 86.2 & 94.8 & 98.0 & 83.6 \\
\quad GRPO speaking-only & 92.4 & 73.6 & 89.7 & 85.5 & 59.1 \\
\quad GRPO window-only & 80.7 & 56.6 & 74.1 & 80.5 & 41.8 \\
\quad \WL & 95.5 & 74.3 & 92.3 & 88.9 & 54.8 \\
\addlinespace[2pt]
\multicolumn{6}{@{}l}{\textit{BayLing}} \\
\quad Base & 81.6 & 65.0 & 90.1 & 66.4 & 77.3 \\
\quad GRPO G16 & 8.1 & 2.0 & 8.7 & 1.7 & 4.3 \\
\quad GRPO speaking-only & 84.8 & 66.7 & 89.1 & 73.0 & 75.3 \\
\quad GRPO window-only & 61.6 & 36.4 & 62.4 & 46.7 & 48.2 \\
\quad \WL & 98.9 & 99.0 & 96.1 & 99.3 & 84.6 \\
\bottomrule
\end{tabular}\hspace{0.5cm}%
\begin{tabular}[t]{@{}lrrrrr@{}}
\toprule
Configuration & ask & deny & repeat & shift & wait \\
\midrule
\multicolumn{6}{@{}l}{\textit{Freeze-Omni}} \\
\quad Base & 2.7 & 17.3 & 1.8 & 2.4 & 34.3 \\
\quad GRPO G16 & 0.0 & 0.0 & 0.0 & 0.0 & 1.4 \\
\quad GRPO speaking-only & 1.3 & 0.0 & 0.0 & 0.0 & 0.6 \\
\quad GRPO window-only & 0.0 & 4.3 & 0.0 & 0.0 & 0.0 \\
\quad \WL & 32.4 & 26.4 & 28.6 & 19.8 & 23.1 \\
\addlinespace[2pt]
\multicolumn{6}{@{}l}{\textit{PersonaPlex}} \\
\quad Base & 97.1 & 94.3 & 95.4 & 95.3 & 87.2 \\
\quad Kyutai GRPO & 99.6 & 97.2 & 97.5 & 97.1 & 88.7 \\
\addlinespace[2pt]
\multicolumn{6}{@{}l}{\textit{Moshika}} \\
\quad Base & 53.4 & 41.6 & 42.5 & 52.0 & 46.4 \\
\quad Kyutai GRPO & 70.0 & 53.1 & 58.7 & 64.3 & 51.1 \\
\bottomrule
\end{tabular}
\caption{Yield rate (\%) by interruption category on the full test set.}
\label{tab:cat}
\end{table}

\paragraph{Zero-advantage windows.} Our GRPO trainers drop rollouts in which the model is not speaking at injection and skip groups with fewer than two valid rollouts or a single reward value; Figure~\ref{fig:diag}a counts both. On Freeze-Omni, 32 of 73 base-policy windows have no valid rollout at all; the zero-advantage groups of Lychee-FD and MiniCPM-o are uniform-reward groups. From the first to the last bin of Figure~\ref{fig:diag}a the share grows from 56\% to 81\% on Lychee-FD and from 77\% to 97\% on Freeze-Omni. Under the window mask the overall shares are 76\% (Lychee-FD), 11\% (MiniCPM-o), 56\% (BayLing) and 82\% (Freeze-Omni), against 72 / 19 / 36 / 91\% for GRPO G16: the mask changes where the credit lands; how often a group is uniform is set by the training trajectory and stays large on three of the four models.

\paragraph{GRPO's criterion and sensitivity.} GRPO G16 crosses $c=0$ on Lychee-FD, MiniCPM-o and BayLing. If the base ROC is proper (concave), a criterion move alone can explain its $d'$ gain on Lychee-FD (+1.8) but not on MiniCPM-o. There, GRPO speaking-only reaches 80.1\% at $B=11.9\%$, above any such ROC through the base point even at the edges of its confidence intervals (such a ROC reaches at most $I=26\%$ at that $B$), so GRPO there also gains sensitivity; GRPO G16 spends part of it on false stops (34.7\%); the speaking-only run does not (11.9\% against the base's 10.0\%). In the direction of Proposition~\ref{prop:bias}, \WL\ moves $c$ toward 0 on three of the four models where it applies, none crossing 0; Freeze-Omni moves away (0.75$\to$0.99). MiniCPM-o's $d'$ of 3.46 rests on $B=0.3\%$ and depends on the correction.
\paragraph{Yield by category and latency.} On Lychee-FD, GRPO G16 yields to 36\% of ``wait'' and 45\% of ``deny'' interruptions, against 92\% and 89\% for \WL\ (Table~\ref{tab:cat}), consistent with lower stop probabilities throughout the window; its gain also comes with slower yields, a median 1.2~s after the interjection against 0.4~s for the base and \WL\ (Table~\ref{tab:main}). \WL\ does not prefer early stops: on MiniCPM-o, whose decisions arrive in 1-s chunks, it yields later than the base (1.36 against 0.92~s).
\paragraph{Where Lychee-FD's leaked credit lands.} On Lychee-FD, whose rollouts begin mid-answer, the listening decisions GRPO G16 credits are those after the model has stopped, when the outcome is already fixed. One candidate cause of the drift toward talking is that in backchannel groups the rollouts that stop wrongly carry a negative advantage onto their post-stop decisions, nearly all to stay silent.

\paragraph{The masks on BayLing.} Speaking-only GRPO leaves BayLing where it was (F1 32.3 to 32.7). Window-only GRPO lifts it to 49.7, but $d'$ stays at $-0.01$: yielding to 51\% of interruptions and 52\% of backchannels; its $c$ moves from $-0.76$ to $-0.03$, which the base-policy push sum of Appendix~\ref{app:theory}, near zero on BayLing, does not predict. On Freeze-Omni it collapses like the other GRPO runs, with no informative group in 123 of 150 steps and 104 valid interruption and 26 valid backchannel items at test.

Table~\ref{tab:sdtfull} gives $d'$ and $c$ (\S\ref{sec:setting}) for every configuration with 95\% bootstrap half-widths (resampling as in Appendix~\ref{app:eval}), and the valid counts; median yield latencies are in Table~\ref{tab:main}. Latency runs from the interjection onset to the stop and is quantised by each interface: 0.4-s control decisions for Lychee-FD, 1-s chunks for MiniCPM-o, 0.8-s chunks for BayLing-Duplex, state-head decisions for Freeze-Omni, and 80-ms frames for the Moshi family (the last voiced frame before a gap of at least 0.8~s). Rates near 0 or 100\% make $d'$ and $c$ depend on the correction; the collapsed Freeze-Omni GRPO runs are listed for completeness only.
\begin{table}[H]
\centering\footnotesize\renewcommand{\arraystretch}{0.88}
\begin{tabular}{@{}lrrrr@{}}
\toprule
Configuration & $d'$ & $c$ & $n_I$ & $n_B$ \\
\midrule
\multicolumn{5}{@{}l}{\textit{Lychee-FD}} \\
\quad Base & 1.04 {\scriptsize$\pm$0.17} & $-$0.53 {\scriptsize$\pm$0.09} & 1369 & 273 \\
\quad GRPO G16 & 2.86 {\scriptsize$\pm$0.37} & 1.14 {\scriptsize$\pm$0.19} & 1500 & 300 \\
\quad GRPO speaking-only & 3.01 {\scriptsize$\pm$0.45} & 0.89 {\scriptsize$\pm$0.22} & 1499 & 300 \\
\quad GRPO window-only & 2.83 {\scriptsize$\pm$0.45} & 0.98 {\scriptsize$\pm$0.22} & 1500 & 300 \\
\quad \WL & 2.87 {\scriptsize$\pm$0.21} & $-$0.30 {\scriptsize$\pm$0.11} & 1475 & 298 \\
\addlinespace[2pt]
\multicolumn{5}{@{}l}{\textit{MiniCPM-o}} \\
\quad Base & 0.30 {\scriptsize$\pm$0.21} & 1.12 {\scriptsize$\pm$0.11} & 1447 & 299 \\
\quad GRPO G16 & 1.80 {\scriptsize$\pm$0.17} & $-$0.51 {\scriptsize$\pm$0.09} & 1440 & 297 \\
\quad GRPO speaking-only & 2.02 {\scriptsize$\pm$0.20} & 0.17 {\scriptsize$\pm$0.10} & 1452 & 295 \\
\quad GRPO window-only & 2.24 {\scriptsize$\pm$0.29} & 0.68 {\scriptsize$\pm$0.14} & 1454 & 292 \\
\quad \WL & 3.46 {\scriptsize$\pm$0.38} & 0.84 {\scriptsize$\pm$0.19} & 1450 & 296 \\
\addlinespace[2pt]
\multicolumn{5}{@{}l}{\textit{BayLing}} \\
\quad Base & $-$0.12 {\scriptsize$\pm$0.17} & $-$0.76 {\scriptsize$\pm$0.09} & 1483 & 298 \\
\quad GRPO G16 & 0.01 {\scriptsize$\pm$0.27} & 1.65 {\scriptsize$\pm$0.14} & 1495 & 299 \\
\quad GRPO speaking-only & $-$0.05 {\scriptsize$\pm$0.18} & $-$0.79 {\scriptsize$\pm$0.09} & 1489 & 299 \\
\quad GRPO window-only & $-$0.01 {\scriptsize$\pm$0.16} & $-$0.03 {\scriptsize$\pm$0.08} & 1397 & 287 \\
\quad \WL & 2.46 {\scriptsize$\pm$0.19} & $-$0.48 {\scriptsize$\pm$0.10} & 1434 & 294 \\
\addlinespace[2pt]
\multicolumn{5}{@{}l}{\textit{Freeze-Omni}} \\
\quad Base & $-$0.88 {\scriptsize$\pm$0.16} & 0.75 {\scriptsize$\pm$0.08} & 1457 & 298 \\
\quad GRPO G16 & $-$0.06 {\scriptsize$\pm$0.33} & 2.62 {\scriptsize$\pm$0.16} & 375 & 104 \\
\quad GRPO speaking-only & 0.14 {\scriptsize$\pm$0.44} & 2.68 {\scriptsize$\pm$0.22} & 779 & 168 \\
\quad GRPO window-only & $-$0.10 {\scriptsize$\pm$0.38} & 2.14 {\scriptsize$\pm$0.19} & 104 & 26 \\
\quad \WL & 0.69 {\scriptsize$\pm$0.21} & 0.99 {\scriptsize$\pm$0.10} & 1448 & 298 \\
\addlinespace[2pt]
\multicolumn{5}{@{}l}{\textit{PersonaPlex}} \\
\quad Base & 1.44 {\scriptsize$\pm$0.18} & $-$0.82 {\scriptsize$\pm$0.09} & 1407 & 274 \\
\quad Kyutai GRPO & 1.39 {\scriptsize$\pm$0.19} & $-$1.05 {\scriptsize$\pm$0.09} & 1411 & 280 \\
\addlinespace[2pt]
\multicolumn{5}{@{}l}{\textit{Moshika}} \\
\quad Base & 0.53 {\scriptsize$\pm$0.19} & 0.34 {\scriptsize$\pm$0.10} & 1200 & 221 \\
\quad Kyutai GRPO & 0.47 {\scriptsize$\pm$0.19} & 0.00 {\scriptsize$\pm$0.10} & 1057 & 214 \\
\bottomrule

\end{tabular}
\caption{Sensitivity $d'$, criterion $c$ (positive: reluctant to stop), and valid interruptions $n_I$ and backchannels $n_B$, for every configuration of Table~\ref{tab:main}.}
\label{tab:sdtfull}
\end{table}

\begin{figure}[H]
\centering
\includegraphics[width=\textwidth]{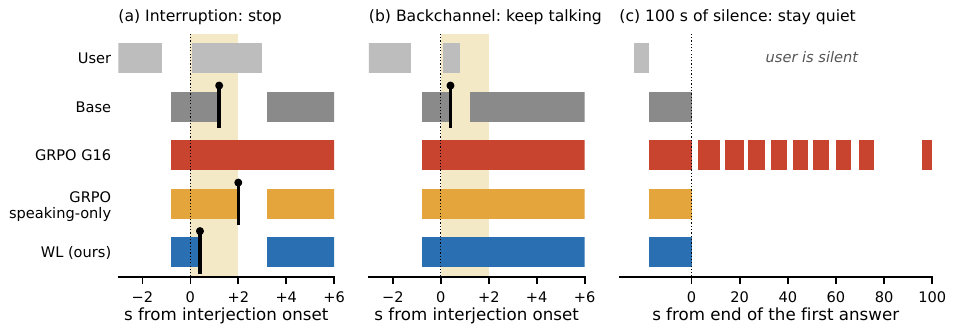}
\caption{Lychee-FD on three test situations (one test item each; rows are the base model and three post-trained configurations). Light grey: user speech; other bars: model speech; black tick: the model's stop decision; shaded band: the 2-s window in which a stop counts. (a) The user asks a follow-up question: the base and \WL\ stop within the window, GRPO G16 never stops, GRPO speaking-only stops at the edge. (b) The user says ``\begin{CJK*}{UTF8}{gbsn}哎是呀\end{CJK*}'' (``oh yes''): the base stops (a false stop), the others keep talking. (c) 100~s of silence after the model's first answer: GRPO G16 re-opens the floor nine times; the others stay quiet.}
\label{fig:timeline}
\end{figure}

\paragraph{Case study.} Figure~\ref{fig:timeline} shows one test item for each of the three situations on Lychee-FD.

\begin{figure}[H]
\centering
\includegraphics[width=\textwidth]{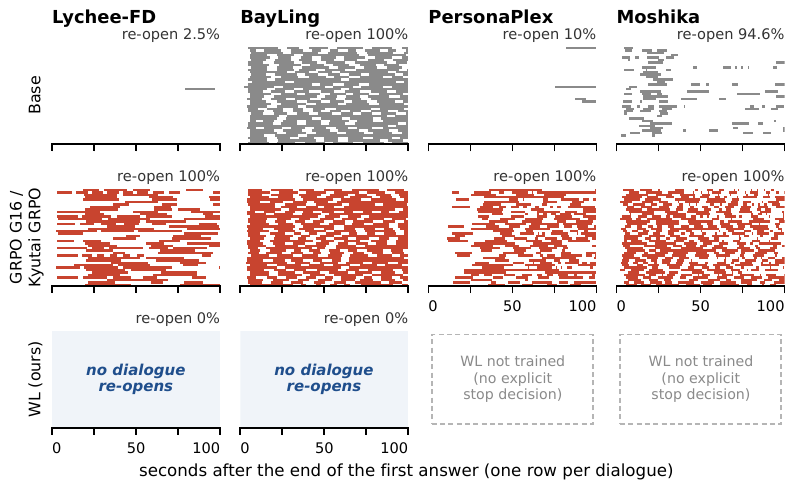}
\caption{The 100-s silence test after the first answer, one row per dialogue, speech as filled bars; an ideal panel has no bars (shaded). GRPO G16 and both Kyutai checkpoints fill the window, as does BayLing's base; \WL\ stays quiet. Freeze-Omni and MiniCPM-o never re-open and are omitted; \WL\ is not trained on PersonaPlex and Moshika (dashed).}
\label{fig:raster}
\end{figure}
\paragraph{What is said into the silence.} Figure~\ref{fig:raster} shows the behaviour behind the last column of Table~\ref{tab:main}; Table~\ref{tab:silence_text} classifies every turn that starts inside the window, from the model's own text channel (Lychee-FD's text tokens, BayLing's chunk text, the Moshi family's inner monologue), by rule and in this order of precedence: \emph{non-verbal} (fewer than four words, including Moshika's wordless audio); \emph{repeat} (a prefix of, or a token-level match of at least 0.8 with, something already said in the dialogue); \emph{imagined reply} (opens with an acknowledgement: ``Yeah'', ``Exactly'', ``That's a great question'', ``You're welcome'', ``I hear you''); \emph{check-in} (at most 30 words offering help or checking on the user: ``anything else?'', ``feel free to ask'', ``let me know if you need anything''); otherwise \emph{new content}. The rules under-count imagined replies.

A re-opening could in principle be a reasonable check on whether the user is still there. None is: no dialogue re-opens with a single check-in, check-ins are at most 22\% of the turns (Lychee-FD GRPO G16, half of them the same sentence recurring across dialogues), and every dialogue that contains one also repeats earlier content, answers a turn that never happened, or continues the answer. The test therefore measures talking into silence, not a check-in.

Most of it follows one of three patterns. \emph{Verbatim loops}: BayLing, before and after any GRPO, re-says one sentence until the window ends (``Some popular drinks include soda, water, and iced tea'', six times in one dialogue), and Lychee-FD GRPO G16 loops in 35\% of dialogues (``If you want to know more about fast food or have other questions, feel free to ask'', nine times). \emph{An imagined interlocutor}: 42\% of Lychee-FD GRPO G16's turns and 38\% of PersonaPlex's Kyutai GRPO turns answer a user turn that never happened (``That's wonderful to hear! You're already off to a great start!''; ``Thanks for talking''), and the model sometimes supplies both sides (``Do you have any other recipes you'd recommend? --- Sure! You could try making a simple pasta dish''; 7\% of Lychee-FD GRPO G16 turns contain a question and its answer). \emph{Continuation}: the Moshi family and BayLing keep extending the list (``Another option is\ldots'', ``And finally\ldots''), BayLing often on an unasked topic (asked about fast food: ``The most popular types of music festivals are rock, pop, and hip hop'').

For evaluation, it matters that the speech is not noise: every Lychee-FD and BayLing turn and 78--88\% of Moshi turns are fluent sentences of eight words or more, so they look like ordinary assistant turns in a transcript; the timing protocol flags them. And the content is the base model's: BayLing's base already loops, the Kyutai checkpoints and Lychee-FD GRPO G16 share the same patterns, and GRPO speaking-only changes BayLing's mix but not the behaviour. Post-training changes whether the model starts talking into silence, not what it says once it has started.

\begin{table}[H]
\centering\footnotesize\renewcommand{\arraystretch}{0.95}
\setlength{\tabcolsep}{2.6pt}
\begin{tabular}{@{}lccccrrrrr@{}}
\toprule
Configuration & Dialogues & Turns & Per & s / & \multicolumn{5}{c}{Share of turns (\%)} \\
\cmidrule(lr){6-10}
 & re-opening & & dlg & turn & Check-in & Repeat & Imagined & New & Non-verbal \\
\midrule
Lychee-FD base & 1/40 & 1 & 1.0 & 17.6 & 0 & 100 & 0 & 0 & 0 \\
Lychee-FD GRPO G16 & 40/40 & 120 & 3.0 & 15.3 & 22 & 15 & 42 & 20 & 0 \\
BayLing base & 40/40 & 224 & 5.6 & 12.3 & 0 & 68 & 0 & 32 & 0 \\
BayLing GRPO G16 & 40/40 & 247 & 6.2 & 11.0 & 0 & 60 & 2 & 38 & 0 \\
BayLing GRPO speaking-only & 40/40 & 236 & 5.9 & 11.3 & 0 & 55 & 0 & 44 & 0 \\
BayLing GRPO window-only & 39/39 & 199 & 5.1 & 14.0 & 0 & 61 & 1 & 38 & 0 \\
PersonaPlex base & 4/40 & 4 & 1.0 & 14.3 & 0 & 0 & 100 & 0 & 0 \\
PersonaPlex Kyutai GRPO & 40/40 & 193 & 4.8 & 8.5 & 5 & 1 & 38 & 50 & 6 \\
Moshika base & 35/37 & 90 & 2.6 & 5.5 & 1 & 1 & 22 & 69 & 7 \\
Moshika Kyutai GRPO & 40/40 & 398 & 9.9 & 5.3 & 6 & 3 & 16 & 62 & 13 \\
\bottomrule
\end{tabular}
\caption{What the models say when they re-open the floor in the 100-s silence test (configurations with at least one re-opening; MiniCPM-o and Freeze-Omni never re-open). Dialogues: re-opening / entered. Turns: speech segments that start inside the window; Per dlg: turns per re-opening dialogue; s/turn: mean seconds per turn. Imagined: imagined reply; New: new content. Categories are assigned by rule in the order non-verbal, repeat, imagined reply, check-in, new content.}
\label{tab:silence_text}
\end{table}

\end{document}